\documentstyle[sprocl]{article}
\input{psfig}
\def\Journal#1#2#3#4{{#1} {\bf #2}, #3 (#4)}
\def\apj{\em Ap. J.}

\def\pasj{\em P. A. S. J.}

\def\al{\alpha}
\def\be{\begin{equation}}
\def\ee{\end{equation}}
\def\bea{\begin{eqnarray}}
\def\eea{\end{eqnarray}}

\begin{document}
\title{ADVECTION-DOMINATED ACCRETION MODEL OF X-RAY NOVA MUSCAE IN OUTBURST}
\author{ A.A. Esin, J.E. McClintock, R. Narayan}
\address{Harvard-Smithsonian Center for Astrophysics, 60 Garden Street,
\\ Cambridge, MA 02138, USA}

\maketitle\abstracts{
We present a model for the high-low state transition of the X-Ray Nova
GS~1124-68 (Nova Muscae 1991) observed by Ginga\,\cite{eb}.  The model
consists of an advection-dominated accretion flow (ADAF) near the
central black hole surrounded by a thin accretion disk.  During the
rise phase of the outburst, as the mass accretion rate increases, the
transition radius between the thin disk and the ADAF moves closer to
the center, until the thin disk extends all the way in to the last
stable orbit.  The transition radius increases again during decline.
We reproduce the basic features of the spectra taken during and after
outburst, and the light curves in the soft and hard X-ray bands.  We
estimate that the accretion rate in Nova Muscae decreased
exponentially with a time scale $\sim 95\,{\rm days}$ during decline.}

\section{Introduction}

A bright X-ray nova GS~1124-68 (Nova Muscae 1991) was discovered by
the Ginga ASM on January 8, 1991.  Subsequent optical
observations{\,\cite{re}} showed that the system is a short period
spectroscopic binary with a mass function of $3.1 M_{\odot}$, which
makes it a strong black-hole candidate.

Like other black hole X-ray novae in outburst, Nova Muscae underwent
large changes in its X-ray luminosity and spectral characteristics
during outburst{\,\cite{eb}} (Figs 1(a), 2).  These changes are
usually described in terms of a succession of spectral states.  The
highest luminosity corresponds to the ``very high'' state,
characterized by a photon index of $\sim 2.5$ in the X-ray range 2-20
keV.  The ``high'' state has a lower bolometric luminosity with
practically no emission above 5-10 keV.  At yet lower luminosity, we
have the ``low'' state where most of the energy comes out in hard
X-rays; the X-ray spectrum is usually well characterized by a
power-law with a photon index $\sim 1.5-1.7$.  Finally, at the lowest
luminosities, there is the ``quiescent
state''{\,\cite{nmy}$^,$\,\cite{nbm}}.

\section{Modeling the State Transitions in Nova Muscae}

We model Nova Muscae as a black hole of mass $M = 6 M_{\odot}$
accreting gas from its companion at a rate $\dot{m}$ (in units of
$\dot{M}_{Edd} = L_{Edd}/(0.1 c^2)$).  From the outer edge at a radius
$r_{out}$ (all radii are in units of $R_{Schw} = 2 G M/c^2$) to a
transition radius, $r_{tr}$, the gas accretes via a cool thin disk
plus a hot corona formed by evaporation of the
disk{\,\cite{nmy}$^,$\,\cite{nbm}}.  Inside the transition radius, the
thin disk is absent and the accretion is via a pure
advection-dominated accretion flow{\,\cite{ny}$^,$\,\cite{a}} (ADAF).

\begin{figure} 
\vbox{\hfil \vbox{\psfig{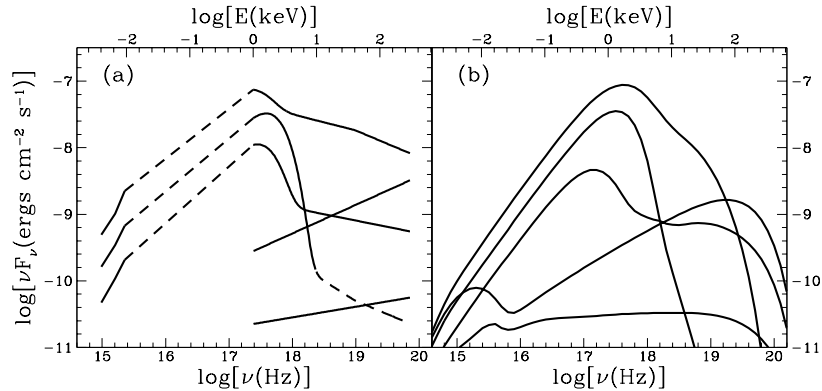}\hfil}}
\vskip -0.2in
\caption{(a) Spectra of GS 1124-68 on days 3 (very high state), 62 
(high state), 130, 197 (low state), and 238 of the outburst (in order
of decreasing $1\,{\rm keV}$ flux).  (b) Corresponding model spectra
with the following parameters: $(\log{\dot{m}},\log{r_{tr}}) =
(0.2,\,0);\ (-0.4,\, 0);\ (-1.0,\,0.5);\ (-1.1,\,3.9);\
(-1.6,\,3.9)$.  A distance of $5\,{\rm kpc}$ was assumed.
\label{fig:spectra}}
\vskip -0.2in
\end{figure}

This two-zone model has been successfully applied to the quiescent
state of two X-ray novae, A0620--00 and V404
Cyg{\,\cite{nmy}$^,$\,\cite{nbm}}.  The model reproduces well the observed
spectra of these systems and provides strong evidence for the presence
of event horizons in the accreting stars.  Here we show that the same
model also explains observations of X-ray novae in outburst.  The
various states arise fairly naturally as $\dot m$ decreases following
the outburst.
 
Above a critical accretion rate $\dot m_{crit}$, it is not possible to
have a pure ADAF without a disk{\,\cite{ny}}.  For such $\dot m$, we
model the flow as a thin disk plus a corona extending all the way from
$\log{r_{out}} = 4.9$ ($80 \%$ of the Roche radius of the primary) to
the last stable orbit at $r=3$.  In our model of Nova Muscae, both the
very high state and the high state correspond to this regime of $\dot
m$.  The distinction between the two is that in the former a large
fraction of the viscous energy released by the disk is dissipated
directly in the corona{\,\cite{hm}}, whereas in the latter the disk
and the corona dissipate only their own viscous energy.  As a result,
in the very high state the system has an active corona which is bright
in hard X-rays while in the high state the corona is very quiet and
radiates very little hard radiation.  The transition between the two
states occurs at $\dot m\approx0.5$.  The top curve in Fig. 2(b) shows
our model spectrum corresponding to the very high state, while the
next curve shows the high state.

Approximately $150\,{\rm days}$ after the peak of the outburst, $\dot
m$ drops below $\dot{m}_{crit} \approx 0.09$, and at this point the
inner regions of the disk evaporate away completely and $r_{tr}$
begins to increase.  This corresponds to the high-low state transition
which was observed in Nova Muscae between 150 and 200 days after the
peak.  With increasing time, the transition radius continues to move
outward until it reaches its quiescent value of $\log(r_{tr}) = 3.9$
(determined from the width of the $H_{\al}$ line in
quiescence\,\cite{nmy}).  At this point the system is in the low state
with $\log{\dot{m}} \approx -1.1$ and the X-ray spectrum is a pure
power law.  At later times, as $\dot m$ decreases further, the system
switches to the quiescent state.

\begin{figure} 
\vbox{\hfil\psfig{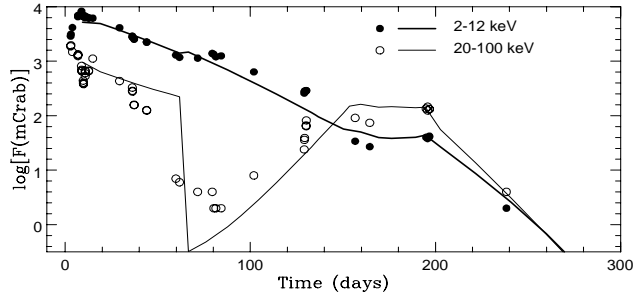}\hfill}
\caption{Filled and open circles trace the observed light curve of 
GS 1124-68 during its 1991 outburst$^1$.  The lines show the
corresponding light curves calculated with our model.
\label{fig:ltcv}}
\vskip -0.2in
\end{figure}

Our model light curve of Nova Muscae is shown in Fig. 2 together with
the data.  For the model we used the scaling: ${\rm Time} = 13
\log{r_{tr}} - 95 \log{(\dot{m}/2.51)}\ {\rm days}$.  The model
satisfactorily reproduces the observed variations in both soft and
hard X-rays.  

In other work, we find that the high-low transition in Cyg X-1 can be
explained through variations in $r_{tr}$, just as in Nova Muscae.

\section*{Acknowledgments} This work was supported in part by
NASA grant NAG 5-2837 and the Smithsonian Scholarly Studies Program.

\end{document}